\documentclass{article}

\usepackage{graphicx}
\usepackage{psfig}
\usepackage{epsfig}
\usepackage[round]{natbib}

\title{
Improving on the empirical covariance matrix using truncated PCA with white noise residuals
 }

\author{Stephen Jewson}
\begin{document}

\author{Stephen Jewson\footnote{\emph{Correspondence address}: Email: \texttt{x@stephenjewson.com}}\\}

\maketitle

\begin{abstract}
The empirical covariance matrix is not necessarily the best estimator
for the population covariance matrix:
we describe a simple method which gives better estimates in two examples.
The method models the covariance matrix using truncated PCA with white noise residuals.
Jack-knife cross-validation is used to find the truncation that maximises the
out-of-sample likelihood score.
\end{abstract}

\section{Introduction}

There are many applications in which it is necessary to estimate
population covariance matrices from sample data.
Our own particular interest is in the statistical modelling of weather data
for the valuation of weather-related insurance contracts~\citep{jewsonbz05},
but there are other uses in fields as diverse as ecology and pattern recognition.
A simple and commonly used estimator for the population covariance matrix is the empirical covariance matrix.
However, there seems to be no reason why this should be the best estimator, and
we present a recipe that we show generates better estimates in two examples.
The recipe is based on PCA. We apply PCA to the sample
data, truncate the series of singular vectors and model the residuals using white noise.
The truncation is then varied and the optimal truncation is chosen as that which
maximises the out-of-sample likelihood in a jack-knife test.
The resulting estimate of the population covariance matrix
is a better estimate than the empirical covariance matrix in the
sense that it gives higher out-of-sample likelihood scores for the sample data.


In section~\ref{pca} we briefly review PCA,
in section~\ref{method} we describe our method for determining the optimal truncation,
in section~\ref{example} we give two examples and
in section~\ref{summary} we summarise.

\section{Principal Component Analysis}
\label{pca}

Consider a matrix of data $X$ with dimensions $s$ by $t$ and rank $r$.
We will think of $s$ and $t$ as representing space and time, but many other
interpretations are possible.
Mathematically speaking, we know that $r \le \mbox{min}(s,t)$. Practically speaking, for any
genuine observed data, we can usually assume that $r=\mbox{min}(s,t)$. This is because
it is infinitely unlikely that there is a linear relation between the columns or the rows
in $X$ (unless one of the columns or rows has deliberately been produced as a linear combination of the others).
Such is the typical nature of real measured data.

The mathematical theory of singular value decomposition states that all matrices can be decomposed
in a certain unique way. Applying this theory to our matrix $X$ gives:

\begin{equation}\label{X=}
 X=E \Lambda P^T
\end{equation}

where $E$ has the dimensions $s$ by $r$, $\Lambda$ has dimensions $r$ by $r$ and $P$ has
dimensions $t$ by $r$. By the singular value decomposition theorem these matrices have the
following properties (\emph{inter alia}):
\begin{itemize}
    \item $E^T E=I$
    \item $P^T P=I$
    \item $\Lambda$ is diagonal
\end{itemize}

PCA is very closely related to eigenvalue decomposition: $E$ contains the eigenvectors of the
covariance matrix $XX^T$, $P$ contains the eigenvectors
of the covariance matrix $X^TX$ and the two covariance matrices have the same eigenvalues,
which are the diagonal terms of  $\Lambda^2$ (we discuss the relations between PCA
and eigenvalue decomposition in a little more detail in~\citet{jewson03x}).

We can write equation~\ref{X=} in terms of the elements of the matrices as:
\begin{equation}\label{x=}
 x_{ij}=\sum_{k=1}^r e_{ik} \lambda_k p_{jk}
\end{equation}

In this form we can see more clearly that we are writing the original data in terms of a sum of $r$
rank 1 matrices, each of which is formed as the product of two vectors and a scalar.
Since we are thinking of the two dimensions as space and time
we can think of the two vectors that make up the $k$'th rank 1 matrix
as being a set of weights in space (a spatial pattern $e_{ik}$) and
a set of weights in time (a time series $p_{jk}$).
The ordering of the rank 1 matrices is arbitrary, but by convention is always taken
with the highest values of $\lambda$ first. This has the consequence that the first of the $r$ matrices
contains the most variance, the second contains the next-most, and so on.
One of the properties of PCA is that the variance accounted for by the first rank 1 matrix
is actually the largest possible
(among all rank 1 matrices, subject to the orthonormality constraints),
and the variance accounted for by the second is the largest possible from the remaining variance.

There are various adaptions of this basic version of PCA.
For instance, the matrix $X$ may be centred and/or standardized prior to deriving the
patterns.

Given equation~\ref{x=} we can consider approximating the data by truncating
the sum to fewer than $r$ of the rank 1 matrices. If we let $r'$ be the number
of matrices retained this gives:

\begin{equation}\label{x-hat=}
 \hat{x}_{ij}=\sum_{k=1}^{r'} e_{ik} \lambda_k p_{jk}
\end{equation}

This truncation may make sense for two reasons. Firstly, the retained patterns
together may account for a large fraction of the total variance, but in only a small
number of patterns. PCA can thus act as an efficient way to represent a large fraction of the information in $X$.
Secondly, the retained patterns are presumably the more accurately estimated patterns,
in a statistical sense. This is useful if the PCA is to be used for simulation or
extrapolation of any kind.

We will now make the restrictive assumption that the data in $X$ is independent in time, dependent in space and
distributed with a multivariate normal distribution.
In this case the spatial patterns show structure while the time series are uncorrelated.
We wish to generate surrogate data that has the same
correlation structure in space as $X$,
and this can be done by replacing the time series
in expression~\ref{x=} with simulated values:

\begin{equation}\label{x-sim=}
 x^{sim}_{ij}=\sum_{k=1}^r e_{ik} \lambda_k p^{sim}_{jk}
\end{equation}

It is easy to show that $x^{sim}$ has the same spatial covariance matrix as the original $x_{ij}$.
However, the rank 1 matrices for high values of $k$ are likely to be very poorly estimated,
and this may be bad for our simulations.
This motivates the idea that we should perhaps truncate the sum and use only the
well estimated patterns in the simulation, up to the $r'$'th.
There are two problems with this, however:
first, that the variance
of the resulting simulated data would be lower than the variance of the observations,
and second that the rank of
the simulated data could be too low (the dimension of the space spanned by the simulated data
could be smaller than the dimension of the space spanned by the sample data).
This might result in simulations which could never explore the space of possible observations
fully, and we find this to be undesirable.
These problems can both be corrected by adding appropriate amounts of white noise as `padding'.

This gives:
\begin{equation}\label{x-sim2=}
 x^{sim}_{ij}=\sum_{k=1}^{r'} p_{ik} \lambda_k q^{sim}_{jk}+\sigma_i \epsilon_{ij}
\end{equation}

where $\epsilon$ is white noise and the $\sigma_i$ are chosen so that the simulations
have the correct variance. The lower $r'$, the greater the $\sigma_i$ have to be to make
up the full variance.

Within this setup the question we wish to ask is: how should the truncation $r'$ be chosen?

\section{Choosing the truncation}
\label{method}

The method we propose for choosing the truncation works as follows.
As the truncation $r'$ is increased, more information about the correlation structure of $X$
is included in the simulations.
But more spurious information is also included because the higher order patterns are less well estimated.
Because of these competing effects the benefit of increasing
$r'$ presumably disappears at some point: we wish to find exactly the value of $r'$ at which
this occurs. To do so we use a jack-knife cross-validation technique: we
test the extent to which a certain truncation is able to represent
data that is outside the sample of data on which the PCA is estimated. This test
allows us to compare different truncations in a fair and honest way, and find
which performs the best.

What cost function should we use for our test? A particular truncation along with the white noise padding
is effectively an estimate of the multivariate distribution of $X$.
This motivates us to use the standard cost function used for the fitting of distributions in classical statistics,
which is the log-likelihood. Given a particular truncation, and the
amplitudes of the supplementary white noise, we can calculate the covariance matrix
of the multivariate distribution.
From this we can calculate the log-likelihood using the standard expression for the density
for the multivariate normal with dimension $p$:
\begin{equation}
 f=\frac{1}{(2\pi)^{\frac{p}{2}} D^\frac{1}{2}} \mbox{exp}\left(-\frac{1}{2}(z-\mu)^T\Sigma^{-1}(z-\mu)\right)
\end{equation}
where
$\Sigma$ is the covariance matrix (size $p$ by $p$),
$D$ is the determinant of the covariance matrix (a single number),
$z$ is a vector length $p$ and
$\mu$ is a vector length $p$.

The log-density is then:
\begin{equation}\label{logf}
 \mbox{log}f=-\frac{1}{2}p\mbox{log}(2\pi)
               -\frac{1}{2}\mbox{log}D
               -\frac{1}{2}(z-\mu)^T\Sigma^{-1}(z-\mu)
\end{equation}

We will refer to the 2nd and 3rd terms of this equation as the `dispersion term' $(-\frac{1}{2}\mbox{log}D)$
and the `standardisation term' $(-\frac{1}{2}(z-\mu)^T\Sigma^{-1}(z-\mu))$.
$D$ is a measure of the dispersion in the multivariate distribution:
for instance, when $p=1$ we have $D=\sigma$. The dispersion term (which has a negative coefficient)
penalizes distributions with a large dispersion.
$(z-\mu)^T\Sigma^{-1}(z-\mu)$ is the `z value' or standardised value of the spatial pattern $z-\mu$,
in the multivariate normal distribution described by $\Sigma$. If $z-\mu$ is very unlikely
in this distribution then this term will be very large. The standardisation term penalizes
the distribution if there are many points with large standardised values.
The distribution which maximises the log-likelihood is a trade-off between these two effects:
the dispersion has to be small, but not so small that the standardised values of the out-of-sample
data is too large.

One aspect of using log-likelihood as a cost function is that it rejects a distribution and covariance matrix completely
if there is even a single observation that could not have come from the distribution. For instance,
if we use truncated PCA without the white noise padding then many of the out-of-sample observations would be
impossible, simply because they come from a higher dimensional space. We consider
this strict rejection of distributions that do not span the space of the observed data to be desirable.

We now summarise our method. For each truncation we run over the data,
missing out each time point in turn, applying PCA to the remaining data,
truncating at the given level, estimating the amplitude of the supplementary white noise,
calculating the covariance matrix for the combination of truncated singular vector series
and white noise,
and calculating the log-likelihood for the missed data. We combine
all the log-likelihoods for a particular truncation to give a single score for that
truncation. We then compare these log-likelihood scores across the different truncations to find
which truncation is the best at predicting the distribution of the out-of-sample data.

\section{Examples}
\label{example}

We now give two simple examples of the method described above.
They are both motivated by our interest in simulating the risk in weather derivative
portfolios, for which we wish to create many thousands of years of surrogate weather data
(see chapter 7 in~\cite{jewsonbz05}).

In both examples we standardise the data in time before we apply PCA.
For the first example $s<t$, while for the second $s>t$.
This alters the nature of the problem significantly, as we will see below.

\subsection{Example 1: UK temperatures}

In our first example we take a matrix $X$ of data consisting of winter average
daily average temperatures for 5 UK locations. There are 44 winters of data and
so $s=5$ and $t=44$. The rank of the data is 5, and is unaffected by the standardisation, which
is only applied in the time dimension.
The space of possible spatial
patterns, which has dimension 5, can be spanned by the 5 spatial singular vectors
if there is no truncation. If there is truncation then this is no longer the case,
and a general spatial pattern could not be represented as a linear combination of
the remaining spatial singular vectors.
The `padding' with white noise solves this problem, as described above.

\begin{figure}[!htb]
  \begin{center}
    \scalebox{0.8}{\includegraphics{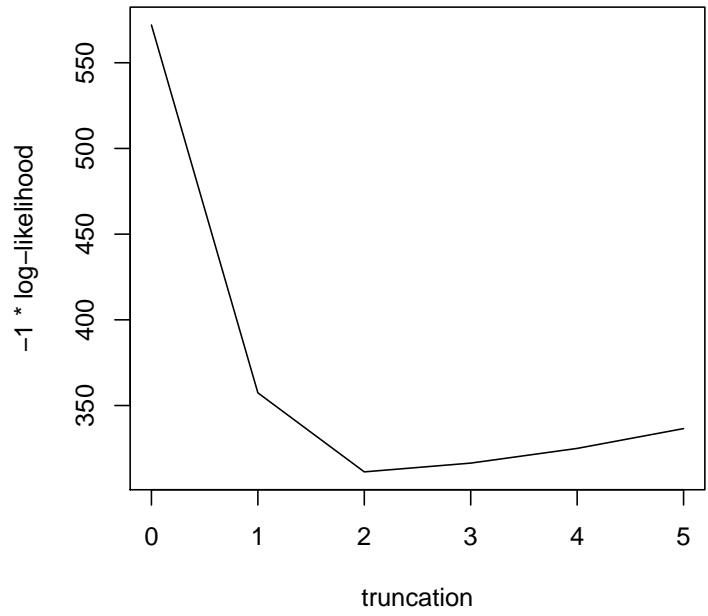}}
  \end{center}
  \caption{
The log-likelihood versus truncation for example 1 described in the text.
     }
  \label{f01}
\end{figure}

Figure~\ref{f01} shows (minus one times) the log-likelihood versus the truncation for this example.
We see that there is a big decrease in the cost function as we move from a purely independent
model to one that uses the first singular vector only:
we conclude that this data is definitely correlated in space.
 There is a much smaller further decrease
when the second singular vector is added, and adding further singular vectors beyond the second
actually increases the cost function.
A truncation to two singular vectors is therefore optimal
in this case.
Truncations of two, three and four all perform better than using the empirical covariance matrix
(which is a truncation of five).
The covariance matrix based on all five singular vectors, and the change in the covariance
matrix caused by truncation to the first two, are shown below. We see that the changes in the
individual covariances are fairly small (perhaps between 1\% and 4\%).

\begin{center}
\begin{tabular}{|c|c|c|c|c|}
  \hline
   46.00 &    42.40 &    37.25 &    41.17 &    40.49 \\
   42.40 &    46.00 &    38.24 &    42.69 &    41.17 \\
   37.25 &    38.24 &    46.00 &    44.04 &    44.46 \\
   41.17 &    42.69 &    44.04 &    46.00 &    44.92 \\
   40.49 &    41.17 &    44.46 &    44.92 &    46.00 \\
  \hline
\end{tabular}
\end{center}

\begin{center}
\begin{tabular}{|c|c|c|c|c|}
  \hline
    0.00 &     1.72 &    -0.35 &     0.39 &    -0.14 \\
    1.72 &     0.00 &     0.16 &    -0.22 &     0.27 \\
   -0.35 &     0.16 &     0.00 &     0.26 &     0.36 \\
    0.39 &    -0.22 &     0.26 &     0.00 &     0.27 \\
   -0.14 &     0.27 &     0.36 &     0.27 &     0.00 \\
  \hline
\end{tabular}
\end{center}

Going further, we can test whether a truncation of two is \emph{significantly}
better than a truncation of one. We will do this using the method we used
in~\citet{hallj05b} in which we consider each individual time point of the data and count
the number of times each of the two methods beats the other. The resulting
test statistic is distributed as a binomial distribution under the null hypothesis
that there is no significant difference between the two truncations.

\begin{figure}[!htb]
  \begin{center}
    \scalebox{0.8}{\includegraphics{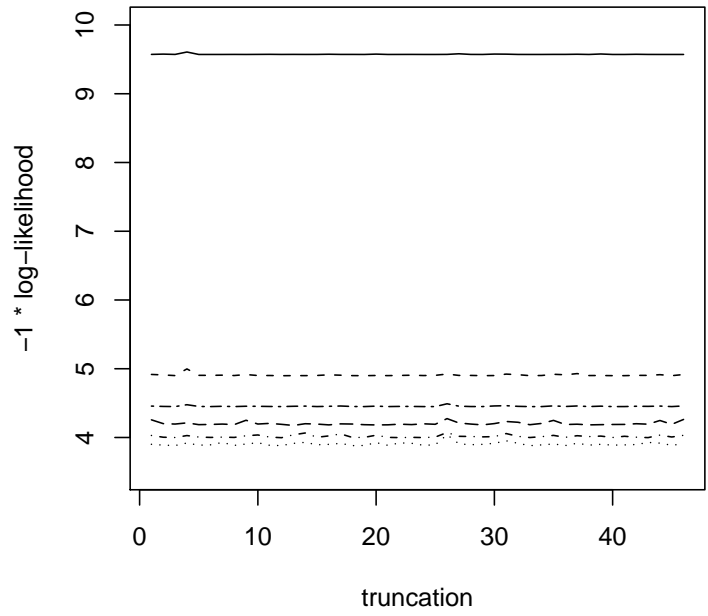}}
  \end{center}
  \caption{
The log-likelihood on a yearly basis for the six truncations used in example 1.
     }
  \label{f02}
\end{figure}

\begin{figure}[!htb]
  \begin{center}
    \scalebox{0.8}{\includegraphics{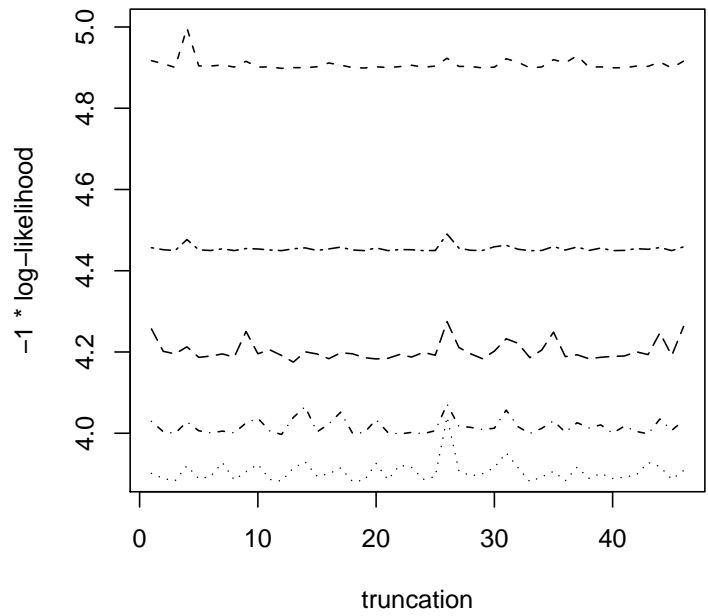}}
  \end{center}
  \caption{
Same as figure~\ref{f02} but with a different scale to clarify the differences between
the curves.
}
  \label{f03}
\end{figure}

The results of this year by year comparison are shown in figures~\ref{f02} and~\ref{f03}.
We see that, for every comparison of adjacent truncations,
one or the other wins \emph{in every year}. We conclude that the ordering of the
results in figure~\ref{f01} is extremely highly significant.

\begin{figure}[!htb]
  \begin{center}
    \scalebox{0.8}{\includegraphics{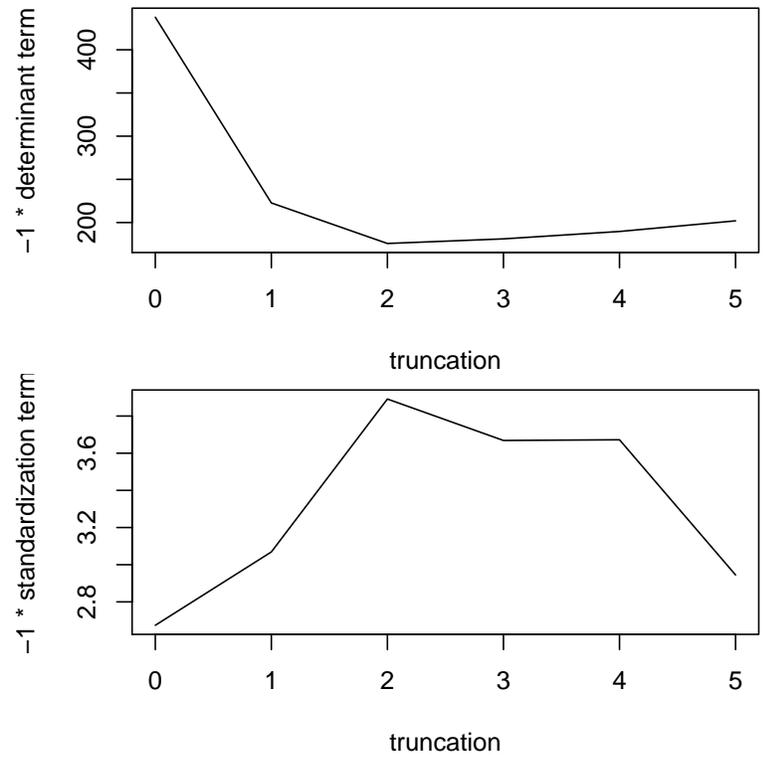}}
  \end{center}
  \caption{
Decomposition of the log-likelihood curve in figure~\ref{f01} into the
determinant and standardization terms. We see that the curve in figure~\ref{f01}
is completely dominated by the determinant term.
    }
  \label{f04}
\end{figure}

We can also try and understand the variations in the log-likelihood score curve shown in figure~\ref{f01}
by breaking the curve down into the determinant and standardization terms in equation~\ref{logf}.
This breakdown is shown in figure~\ref{f04}. We see that, in this case, the shape of the log-likelihood
score curve is fixed by the determinant term. Had we known this in advance we
could have found the optimum truncation by simply calculating the determinant as a function of
truncation. This is a simple in-sample calculation, and much less complex than the full
cross-validation calculation. We suspect that it may always be the case
that the determinant term dominates when $s<t$, and this possibility seems to merit further investigation.
We also suspect that the dominance of the determinant term explains why the breakdown by year
gives such clear results.

With some trepidation we now attempt to explain the behaviour of the determinant and
standardisation curves. The standardisation curve seems to be the easier of the two
to understand. For all 6 truncations this term is very small: this means that all of the out-of-sample
spatial patterns are quite consistent with the fitted distribution. This is presumably because
the out-of-sample patterns live in a 5 dimensional space, and the fitted distributions
have significant variance in all of these dimensions.
The determinant curve is a little harder to understand. As the truncation increases
it shows a decrease and then an increase.
The decrease seems to be because as the truncation is increased the degree of specialisation
of the model increases. The subsequent increase is presumably because of sampling error
on the higher singular vectors.

\subsection{Example 2: US temperatures}

In our second example we take a matrix $X$ of data consisting of winter average
daily average temperatures for 308 US locations. There are 54 winters of data and
so $s=308$ and $t=54$. The rank of the data is 53 because of the temporal standardisation.
Because $s>t$ we are now
in a situation where the space of possible spatial patterns, which has dimension
308, cannot be spanned by the spatial singular vectors, of which there are only 53.
Truncation and the white noise padding are therefore essential: this is a case
where it seems that we are \emph{guaranteed} to find a better estimate of the covariance
matrix than that given by the empirical covariance matrix, because the empirical
covariance matrix will immediately fail. In fact, the simple example of a purely independent model
(a full-rank diagonal covariance matrix) will always beat the empirical covariance matrix.

\begin{figure}[!htb]
  \begin{center}
    \scalebox{0.8}{\includegraphics{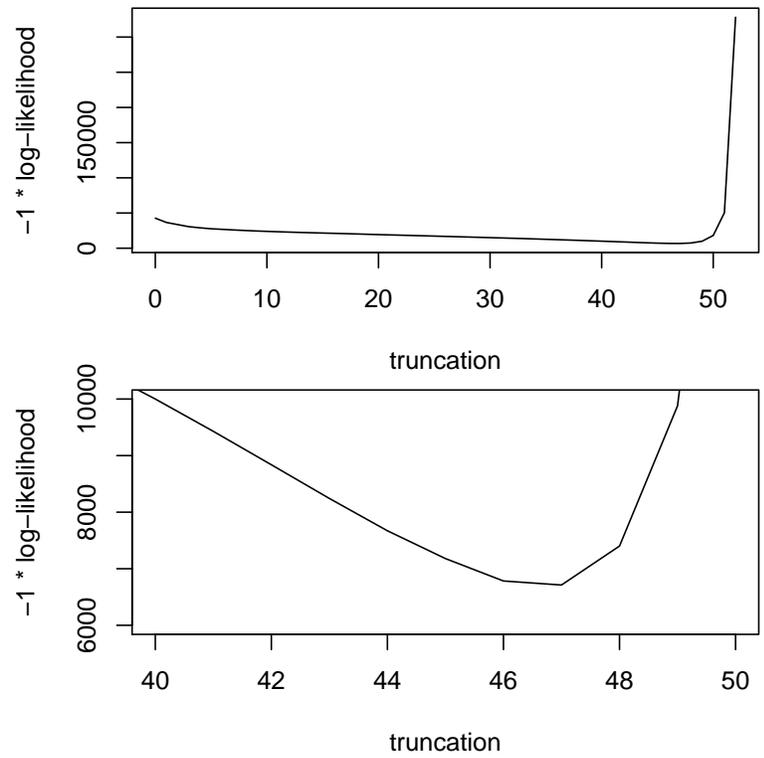}}
  \end{center}
  \caption{
The log-likelihood versus truncation for example 2 described in the text,
with two different vertical and horizontal scales.
    }
  \label{f05}
\end{figure}

The likelihood score versus truncation is shown in figure~\ref{f05}.
We can only evaluate the likelihood score up to a truncation of 52. This is because
the rank of the data is 53, and so the truncation of 53, which has no white noise
padding, gives a correlation matrix that cannot be inverted.

We see that the log-likelihood gradually reduces as the truncation is increased, up to
a truncation of 47. It then rapidly increases to very large values between 47 and 52.
47 is thus the optimum truncation.

\begin{figure}[!htb]
  \begin{center}
    \scalebox{0.8}{\includegraphics{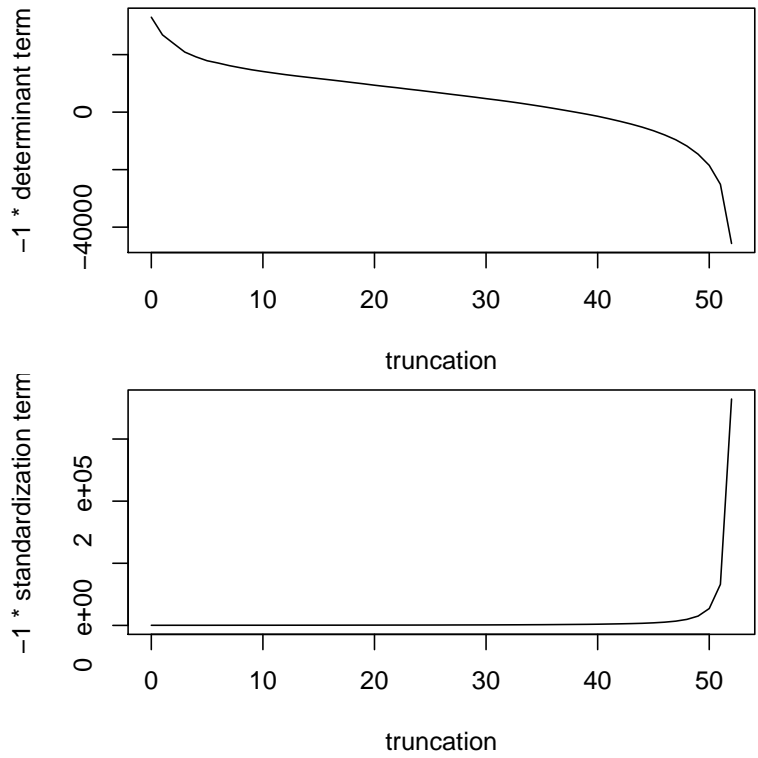}}
  \end{center}
  \caption{
Decomposition of the log-likelihood curve in figure~\ref{f05} into the
determinant and standardization terms. In this case the curve in figure~\ref{f05}
is not dominated by either term, and the minimum in the curve in figure~\ref{f05}
arises from interplay between these two terms.
    }
  \label{f06}
\end{figure}

In figure~\ref{f06} we decompose the log-likelihood curve into determinant
and standardization terms. In this case we see that it is the interplay of these two
terms that fixes the minimum, and it would not be possible to determine the minimum
using the determinant curve alone (which is monotonic).

Again, with some trepidation, we attempt to explain the shapes of these two curves.
The determinant curve decreases as the truncation increases: we think this is
because adding more singular vectors, at the expense of white noise variance,
makes the multivariate distribution more specific i.e. it concentrates the
variance into fewer dimensions. Ultimately, for a truncation of 53, there is only
non-zero variance in 53 of the 308 dimensions (and the correlation matrix
is no longer invertible). The standardisation term gradually increases
as a result of this specialisation. Then, as the truncation approaches
53, the variance in the other dimensions becomes very small, and the probability
of some of the out of sample patterns, which come from a 308 dimensional space,
becomes very low. At this point the standardisation term becomes very large.
We think that this tradeoff between the determinant term and the standardisation
term is likely to occur whenever $s>t$.

\section{Summary}
\label{summary}

We have investigated a simple approach for making a better estimate of the population covariance
matrix than that given by the empirical covariance matrix.
The method is based on truncated PCA with white noise residuals.
The question of how to truncate PCA has been addressed before, but we introduce
a simple new method based on a very straightforward reasoning: we want to choose the truncation
so that we maximise the likelihood of out-of-sample data. Finding the best truncation
under this definition of optimum is relatively easy. We give two examples, and in both
cases we find better estimates of the population covariance matrix than that given by
the empirical covariance matrix (where \emph{better} is defined as giving higher
out-of-sample likelihood scores).

Based on the results from our examples we conclude that using
the empirical covariance matrix for statistical modelling
may not be a very good
idea since the higher order singular vectors tend to be poorly estimated and thus decrease the
out-of-sample likelihood.
In the $s>t$ case there is the additional problem that the empirical covariance matrix does not
describe a space large enough to contain the observations.
Optimal truncation with white noise `padding' solves both these problems,
and thus may give better modelling results.

In some cases, such as the two examples we have used in this study, one of the dimensions of the
sample data is a genuine spatial dimension. In this case it may be possible to do even better by
modelling the residuals using `red' noise, rather than just white noise. Testing this idea is next.
It would also be interesting to compare our method with other possible methods for improving
the estimate of the covariance matrix, such as linear combinations of the empirical covariance
matrix with an independent model.

\section{Acknowledgements}

The author would like to think Dag Lohmann, Sergio Pezzuli and
Christine Ziehmann for interesting discussions on this topic.

\section{Legal statement}

SJ was employed by RMS at the time that this article was written.

However, neither the research behind this article nor the writing
of this article were in the course of his employment, (where 'in
the course of their employment' is within the meaning of the
Copyright, Designs and Patents Act 1988, Section 11), nor were
they in the course of his normal duties, or in the course of
duties falling outside his normal duties but specifically assigned
to him (where 'in the course of his normal duties' and 'in the
course of duties falling outside his normal duties' are within the
meanings of the Patents Act 1977, Section 39). Furthermore the
article does not contain any proprietary information or trade
secrets of RMS. As a result, the author is the owner of all the
intellectual property rights (including, but not limited to,
copyright, moral rights, design rights and rights to inventions)
associated with and arising from this article. The author reserves
all these rights. No-one may reproduce, store or transmit, in any
form or by any means, any part of this article without the
author's prior written permission. The moral rights of the author
have been asserted.

The contents of this article reflect the author's personal
opinions at the point in time at which this article was submitted
for publication. However, by the very nature of ongoing research,
they do not necessarily reflect the author's current opinions. In
addition, they do not necessarily reflect the opinions of the
author's employers.

\bibliography{pca}

\newpage

\newpage

\newpage

\newpage

\newpage

\newpage

\end{document}